\documentclass{svproc}
\usepackage{url}

\usepackage[utf8]{inputenc}

\usepackage{xcolor}

\usepackage{physics, amsmath, amssymb}
\usepackage{graphicx}
\usepackage{multirow}
\usepackage{booktabs}

\usepackage{hyperref}
\hypersetup{
    colorlinks = true,
    citecolor = {blue},
    linkcolor = {blue},
    urlcolor = {blue},
}

\usepackage{cleveref}

\usepackage{etoc}
\etocsetnexttocdepth{subsection}

\begin{document}
\mainmatter              
%
\title{The distance backbone of directed networks}
\titlerunning{The distance backbone of directed networks}  
%
\author{%
    Felipe Xavier Costa\inst{1,2,3}$^{,\dagger}$
    \and
    Rion Brattig Correia\inst{1,3}$^{,\dagger}$
    \and
    Luis M. Rocha\inst{1, 3}$^{,\star}$
}

\authorrunning{Costa \textit{et al.}} 
%
%
\institute{%
    Systems Science and Industrial Engineering Department, Binghamton University (State University of New York), Binghamton NY 13902, USA
    \and
    Department of Physics, State University of New York at Albany, Albany NY 12222, USA
    \and
    Instituto Gulbenkian de Ciência, Oeiras 2780-156, Portugal \\
    $^{\dagger}$equal contribution;
    \email{$^{\star}$rocha@binghamton.edu}
}

\maketitle

\begin{abstract}
    In weighted graphs the shortest path between two nodes is often reached through an indirect path, out of all possible connections,
    leading to structural redundancies which 
    play key roles in the dynamics and evolution of complex networks.
    We have previously developed a parameter-free, algebraically-principled methodology to uncover such redundancy and reveal the distance backbone of weighted graphs, which has been shown to be important in transmission dynamics, inference of important paths, and quantifying the robustness of networks.
    However, the method was developed for undirected graphs.
    Here we expand this methodology to weighted directed graphs and study the redundancy and robustness found in nine networks ranging from social, biomedical, and technical systems.
    We found that similarly to undirected graphs, directed graphs in general also contain a large amount of redundancy, as measured by the size of their (directed) distance backbone.
    %
    %
    Our methodology adds an additional tool to the principled sparsification of complex networks and the measure of their robustness.
\end{abstract}

\section{Introduction}
\label{sec:intro}

Networks are a canonical method to model complex multivariate interactions and have been proven useful in the study of a variety of problems, such as social interaction and human mobility to predicting epidemic spreading \cite{Colizza:2006,Karsai:2011,Mistry:2021} and modeling biochemical networks to predict the onset of diseases \cite{Gates:2021,Correia:2022:meionav}.
This modeling approach allows for a shift from the traditional scientific focus on the (reductionist) study of things (e.g., animals or proteins), to the study of system-wide interactions among these things, such as friendships among animals, or bonding among proteins.
In network science, typically, these multivariate interactions are represented as edges that connect variables as nodes in a graph.
In addition, networks built to represent real-world complex systems often denote variable interaction with a weight that is proportional to the strength of interaction between nodes, such as a proximity (similarity) or a distance (dissimilarity).
For instance, edge weights can represent the probability of interaction between genes \cite{Correia:2022:meionav}, similarity between concepts in a knowledge space \cite{Correia:2016}, or a measure of how much time two individuals spent together in close vicinity \cite{Correia:2022:epibackbone}.
In its simplest form, edge weights are non-directed, meaning interactions between nodes are symmetric.
This is especially the case when distance and shortest paths between nodes are relevant for analysis---e.g. inferring the likelihood that a person infects another in a population under epidemic spread---because distance measures are by default symmetric (in addition to being non-negative and anti-reflexive \cite{Simas:2015}).

Redundancy is considered a fundamental aspect in the evolution of complex systems \cite{Conrad:1990}. Distinct aspects of the phenomenon have been shown to greatly contribute to our understanding of network dynamics, controllability, and robustness \cite{Gates:2016,Gates:2019,Simas:2021}.
In particular, we have shown that most networks where edges represent distance (or dissimilarity) contain large amounts of topological redundancy in computing shortest paths, which can be identified through our algebraically-principled and parameter-free \textit{distance backbone} \cite{Simas:2021}.
This means our method differs from other backbones by requiring no tunning parameter, null model comparisons, or Monte Carlo approximations.
However, even though distance is typically considered to be symmetric \cite{galvin1991distance}, many real-world complex systems are best modeled by \textit{directed, weighted graph}s.
Indeed, asymmetric interactions have been to shown to be important in a variety of domains, ranging from unreciprocated friendships \cite{Ball:2013}, food-webs and host-parasite ecological networks \cite{Ings:2009}, to designing smarter urban traffic and cities \cite{Salgado:2021,Alves:2021}.

Here, our main contribution is the extension of the distance backbone methodology to \emph{directed} weighted graphs.
Specifically, we build upon the concepts of transitive and distance closure for undirected weighted graphs \cite{Simas:2015} to identify a subgraph whose edges do not break a generalized triangle inequality and which are sufficient to compute all shortest \emph{directed} paths.
In other words, we obtain a \emph{directed distance backbone} that preserves the distribution of shortest paths in directed weighted graphs.
This in turn allows us to quantify both the structural redundancy of such networks and their robustness to random attacks.
Real-world examples also show preliminary results that
having directed edges yields a larger distance backbone than it does for undirected graphs.

\section{Closures in complex networks}
\label{sec:methods}

In social networks, indirect associations are often exemplified as ``the friend of my friend is also my friend''.
These indirect associations can be described in a graph $G(X)$, defined on the set of nodes $X$, in terms of the transitive and distance closures.
Transitive closures assume edge weights to measure a similarity while distance closures assume weights to be a dissimilarity between nodes \cite{Simas:2015}.
The formalism for closures in weighted undirected networks has been introduced in Simas \emph{et. al} \cite{Simas:2021}.
We revise this mathematical construction in this section and, in \cref{sec:results}, we relax the symmetry condition previously considered while showing that the formalism of closures in complex networks is applicable to both undirected and directed networks.

\subsection{Transitive closure}
\label{sec:TransC}

The strength of interactions between the nodes $x_i \in X$ can be measured by a proximity graph, $P(X)$.
This is a reflexive network with edges weights $p_{ij} \in [0, 1]$, a continuous range of values, with $p_{ii} = 1$.
Transitivity is computed via the composition of generalized, weighted logical operators.
These are extensions of the binary logic operators, derived from probabilistic metric spaces and fuzzy logic, and are called triangular norms and conorms \cite{Klir:1995,Simas:2015,Simas:2021}.

A triangular norm (t-norm) is a generalized logical conjunction given by the operation $\wedge: [0, 1]\cross[0,1] \to [0, 1]$.
It satisfies the properties of
    commutativity ($p\wedge q = q\wedge p$),
    associativity ($p\wedge(q\wedge w) = (p\wedge q)\wedge w$),
    monotonicity ($p\wedge q \leq w\wedge v$ implies $p \leq w$ and $q\leq v$ ),
    and having $1$ as its identity element ($p\wedge 1 = p$).
Similarly, a triangular conorm (t-conorm) is a generalized logical disjunction given by the operation $\vee: [0, 1]\cross[0,1] \to [0, 1]$.
It is also commutative, associative, monotonic, but has $0$ as its identity element ($p\vee 0 = p$).
Combining them gives us the compositions of $P$ with itself as
\begin{equation}
    P^{\eta} = P\circ P^{\eta-1} \iff p_{ij}^{(\eta)} = \underset{k}{\vee}\qty(p_{ik} \wedge p_{kj}^{(\eta-1)}),
    \label{eq:PComp}
\end{equation}
considering $\eta\in\mathbb{Z}\geq 2$ and $P^1 = P$.
This leads to the transitive closure of $P(X)$ given by
\begin{equation}
    P^T(X) = \displaystyle\bigcup_{\eta=1}^{\kappa} P^{\eta} \iff p_{ij}^T = p_{ij} \vee p_{kj}^{(2)} \vee \cdots \vee p_{ij}^{(\kappa-1)}\vee p_{ij}^{(\kappa)}.
    \label{eq:PClos}
\end{equation}
For general t-norms and t-conorms the closure is reached as $\kappa \to \infty$. But with proximity graphs, as long as  $\land \equiv \min$, the closure $P^T(X)$ converges for a finite $\kappa$ no larger than the graph diameter \cite{Klir:1995,Simas:2015}.
The adjacency matrix $P^\eta(X)$ measures the proximity for paths of size $\eta$, while the transitive closure $P^T(X)$ accounts for the strongest proximity for paths \emph{up to} size $\kappa$.

We say that a proximity graph is transitive with respect to the algebraic structure $([0, 1], \vee, \wedge)$ if for every weighted edge $p_{ij}$ in the graph we have: 
\begin{equation}
    p_{ij} \geq \underset{k}{\vee}(p_{ik} \wedge p_{kj})
    \label{eq:PTrans}
\end{equation}
\noindent
for any node $x_k \in X$.
By construction, all edges of $P^T(X)$ obey this generalized transitivity constraint, while only a subset of edges of $P(X)$ typically do. 
%
%
In the context of the generalized transitivity criterion given by \cref{eq:PTrans}, fully transitive graphs denote a similarity multivariate relation, whereas graphs that break transitivity for at least one edge denote a proximity relation \cite{Klir:1995}.

For connected, undirected graphs, this leads to a closure where $p^T_{ij}>0$ for all $x_i$ and $x_j$ in $X$, i.e. a complete or fully connected graph.
Unfortunately, this does not generalize for directed graphs, where there can be nodes that only have outwards connections, and therefore can never be reached from other nodes.

\subsection{Distance closure}
\label{sec:DistC}

In network science, we often need to compute shortest paths on graphs to infer the (direct and indirect) influence of variables on one another. This requires casting the network as a distance (or dissimilarity) graphs, $D(X)$ on the set of node variables $X$. 
These graphs have non-negative weights, i.e. adjacency matrix elements $d_{ij} \in [0, \infty)$, and are anti-reflexive: $d_{ii} = 0$.
They are also isomorphic to proximity graphs \cite{Simas:2015} via a strictly monotonic decreasing map $\varphi: [0, 1] \to [0, \infty)$ constrained by:
\begin{equation}
    \underset{k}{f}\{ g( \varphi(p_{ik}), \varphi(p_{kj}) ) \} = \varphi(\underset{k}{\vee}(p_{ik} \wedge p_{kj})) \quad \forall x_i, x_j, x_k \in X,
    \label{eq:Isomorphism}
\end{equation}
where $f$ and $g$ are isomorphic operations to $\wedge$ and $\vee$, respectively, in the sense that they are associative, commutative, monotonic, and having identity elements given by $\varphi(0) \to \infty$ for $f$ and $\varphi(1) = 0$ for $g$.
Due to this construction, $g$ and $f$ are named triangular distance norm (td-norm) and conorm (td-conorm), respectively \cite{Simas:2015}.

Though an infinite number of maps satisfy the isomorphism, the simplest, which we use here unless otherwise noted, is the familiar distance function:
\begin{equation}
    d_{ij} = \varphi(p_{ij}) = \frac1{p_{ij}} -1,
    \label{eq:map}
\end{equation}
\noindent
that easily converts between proximity $P(X)$ and distance $D(X)$ graphs.
In addition to being non-negative and anti-reflexive, distance measures are typically symmetrical, and if transitive, are also known as metric \cite{galvin1991distance}.

\Cref{eq:Isomorphism} allows us to study transitivity of distance graphs by establishing an isomorphism with transitive closures of proximity graphs. 
Thus, the distance closure $D^T(X)$ is obtained via compositions of $f$ and $g$:
\begin{equation}
    d_{ij}^{(\eta)} = \underset{k}{f} g\qty( d_{ik}, d_{kj}^{(\eta-1)} ) \qq{\&}
    d_{ij}^{T} = f\qty( d_{ij},  d_{ij}^{(2)}, \cdots , d_{ij}^{(\kappa-1)}, d_{ij}^{(\kappa)}),
    \label{eq:DClos}
\end{equation}
where, because of the isomorphism, $\kappa$ is the same as for the transitive closure (\cref{eq:PClos}).
The adjacency matrix $D^\eta(X)$ measures the shortest distance for paths including $\eta$ connections, while the distance closure $D^T(X)$ accounts for the shortest path length \emph{up to} $\kappa$ links.
For distance graphs, the transitivity criterion is defined by each algebraic structure $([0, \infty), f, g)$: 
\begin{equation}
    d_{ij} \leq \underset{k}{f} g( d_{ik}, d_{kj} ) \quad \forall x_i, x_j, x_k \in X.
    \label{eq:DTrans}
\end{equation}
\noindent
The distance closure $D^T(X)$ is transitive by construction, but generally only a subset of edges $D(X)$ obey \cref{eq:DTrans}.

\subsection{Shortest-path, metric and ultrametric closures}
\label{sec:MetricUltraClos}

The general transitive and distance closures of \cref{sec:TransC,sec:DistC} yield a number of well-known cases used in network science \cite{Simas:2015,Simas:2021}.
When $f \equiv min$ (or $\vee \equiv max$ in proximity graphs), we have the large class of \textit{shortest-path closures}, $D^{T, g}(X)$, for any distance function $g$ (or $\wedge$ in proximity graphs), as the closure selects the minimum path with length given by $g$. 
This leads to a \textit{generalized triangle inequality} \cite{Simas:2021} as a transitivity criterion:
\begin{equation}
    d_{ij} \leq  g( d_{ik}, d_{kj} ) \quad \forall x_i, x_j, x_k \in X.
    \label{eq:SPInequality}
\end{equation}

For instance, when $g \equiv +$, we obtain the familiar \textit{metric closure}, $D^{T, m}(X)$, where the length of the path is obtained by summing the distance edge weights.
In this case, the transitivity criterion becomes the traditional triangle inequality:
\begin{equation}
    d_{ij} \leq d_{ik} + d_{kj} \quad \forall x_i, x_j, x_k \in X.
    \label{eq:TrigIneq}
\end{equation}
\noindent
Similarly, when $g \equiv \max$, we instead obtain the \textit{ultrametric closure}, $D^{T, u}$, where the length of the path is obtained by the maximum distance weight in path (the weakest link), and the transitivity criterion is:
\begin{equation}
    d_{ij} \leq \max (d_{ik}, d_{kj}) \quad \forall x_i, x_j, x_k \in X,
    \label{eq:UltraTrigIneq}
\end{equation}

Many other shortest-path distance closures---and thus different path length measures and transitivity criteria---can be usefully employed in network science \cite{Simas:2015}.
Here we exemplify the approach with these two well-known cases because the metric closure is the most common way to compute shortest path on weighted graphs, and the ultra-metric closure is the lower bound of distance closures \cite{Simas:2021}.

\subsection{Distance backbone subgraph}
\label{sec:Backbone}

The \textit{distance backbone} $B^g(X)$ of a distance graph $D(X)$ is the invariant subgraph under a shortest-path distance closure $D^{T, g}(X)$ with $f \equiv \min$ and some $g$ \cite{Simas:2021}.
It is sufficient to compute all shortest paths in $D(X)$ given a path length measure $g$.
The distance backbone is invariant because its edges are the ones that obey the generalized triangle inequality (\cref{eq:SPInequality}) and are thus called \textit{triangular} edges.
That is, the distance backbone is defined by edges that have the same weight in the shortest-path closure:
\begin{align}
    b^g_{ij} = \begin{cases} d_{ij}, & \text{if } d_{ij} = d^{T,g}_{ij} \\[1ex] \infty, & \text{if } d_{ij} > d^{T,g}_{ij}\end{cases}, \quad\forall x_i, x_j \in X,
    \label{eq:backbone}
\end{align}
where $d^{T,g}_{ij}$ are the adjacency matrix weights of the distance closure graph $D^{T,g}(X)$. 
The edges that break the generalized triangle inequality are called \textit{semi-triangular} and are not on the backbone, i.e $b^g_{ij} = \infty$. 
If (and only if) an edge between $x_i$ and $x_j$ is semi-triangular (i.e., not present on the backbone), there exists a shorter indirect path (i.e., which is present on the backbone) connecting them via some $x_k$ \cite{Simas:2021}.

The metric ($g \equiv +$) and ultrametric ($g \equiv \max$) backbones of distance graph $D(X)$ are denoted by $B^m(X)$ and $B^u(X)$, respectively.
Similarly, edges on these backbones are called metric and ultrametric, while those off are known as semi-metric and semi-ultrametric, respectively \cite{Simas:2021}.

\section{Directed distance backbone}
\label{sec:results}


Here we extend the concept of distance backbone by relaxing the symmetry constraint of distance functions, thus considering distance graphs $D(X)$ where $d_{ij} \neq d_{ji}$, or directed distance graphs.
As summarized above, distance backbones exist when enforcing a generalized triangle inequality (\cref{eq:SPInequality}) as a transitive closure criterion.
This is the same as computing all shortest paths of $D(X)$ using a measure of path length determined by $g$.

Computation of the all pairs shortest path problem (APSP) for undirected
weighted graphs with $g \equiv +$ is straightforward using the Dijkstra algorithm \cite{Brandes:2005} (though it can also be computed with the distance product directly via \cref{eq:PClos,eq:DClos} \cite{Simas:2015,zwick2002all}).
Since all shortest-path distance closures are based on setting $f \equiv \min $ in \cref{eq:DClos,eq:DTrans}, they can also be computed as a APSP problem by adjusting the chosen algorithm with a different path length measure for each $g$ used, such as $g \equiv \max$ for the ultrametric backbone \cite{Simas:2021}.

We also know that the standard triangle inequality (\cref{eq:TrigIneq}) enforced by $g \equiv +$ is valid for directed distances 
\cite{Lawvere:1983}.
This way, the APSP of directed distance graphs based on this transitivity criterion can also be computed via the Dijkstra algorithm \cite{Brandes:2005} or the distance product \cite{zwick2002all}.
%
%
%
Indeed, the methodology of closures in complex networks is found to be applicable to both undirected and directed weighted graphs. The latter is shown in the real world examples of \cref{sec:Experiments}.

\subsection{Redundancy and robustness}

The fraction of edges in the backbone
\begin{equation}
    \tau^g(X) = \frac{\abs{B^g(X)}}{\abs{D(X)}} = \frac{\abs{\qty{d_{ij}: d_{ij} = d_{ij}^{T, g} }}}{\abs{\qty{d_{ij}}}} \,\,\forall_{x_i, x_j \in X: i\neq j} 
\end{equation}
measures the proportion of triangular (or topologically invariant) edges, while its complement $\sigma(X) = 1 - \tau(X)$ quantifies the proportion of semi-triangular edges.
The latter measures the structural redundancy of complex networks given a specific transitivity criterion (eq. \ref{eq:SPInequality}).
That is, the edges that are redundant for shortest-path computation given the path length measure $g$ chosen.
Note that due to the introducing of directionality, now $\tau^g$ must be computed for \emph{all} entries of the adjacency matrix, and not just for the upper or lower diagonal as previously done for the undirected case \cite{Simas:2021}.

If a network has a small backbone (small $\tau^g$), most of its edges are semi-triangular and do not affect the shortest path distribution. 
This means that \emph{random} attacks would most likely not interfere with the backbone itself, a robustness %
    \footnote{
        A finer characterization of robustness in terms of edge properties \cite{Simas:2012} in the case of directed graphs is left for future work.
    } %
that can be inferred from the measure of topological redundancy $\sigma^g(X) = 1 - \tau^g(X)$.

\section{Experimental Analysis}
\label{sec:Experiments}

Now we investigate the backbone of nine real-world networks pertaining to three distinct domains: biomedical, social, and man-made technological systems.
Here we discuss in more detail the backbones of
    a giraffe social network \cite{Bashaw:2007},
    the U.S airport transportation system \cite{Serrano:2009,Simas:2021},
    and the bike-sharing system of the City of London \cite{MunozMendez:2018}.
Additional details for this and the remaining networks can be found in the accompanying digital \href{https://casci.binghamton.edu/publications/CN22-dbdn.php}{supplemental material}. 
Descriptive data for each directed weighted graph, and the size of their respective metric and ultrametric backbone are shown in \cref{tab:summary-results}.

\begin{table}
    \footnotesize
    \centering
    \caption{
        Topological invariance of weighted directed graphs modeling real-world systems.
        The number of nodes $\abs{X}$ and edges $\abs{d_{i\neq j}}$ are used to compute the network density $\delta$.
        The relative size of the metric ($\tau^m$) and ultrametric ($\tau^u$) backbones are presented as percentages.
        %
    }
    \tabcolsep=0.11cm
    \begin{tabular}{ll|rrr|c|c|c}
        \toprule
        & Network & $\abs{X}$ & $\abs{d_{i\neq j}}$ & $\delta$ & $\tau^m$ & $\tau^u$ & $\tau^u/\tau^m$ \\
        \midrule
        \multirow{3}{*}{Biomedical} & Co-morbidity risk & 95 & 8,930 & 1.0 & 47.44 & 2.17 & 4.57 \\
        & Drug interaction & 412 & 2,966 & 1.75e-2 & 59.00 & 40.49 & 68.63 \\
        & Species-Species inter. & 10,578 & 18,529 & 1.66e-4 & 99.47 & 99.46 & 99.99 \\ 
        \midrule
        \multirow{2}{*}{Social} & Giraffe socialization & 6 & 30 & 1.0 & 76.67 & 30.00 & 39.13 \\
        & Telephone calls & 322 & 609 & 5.89e-3 & 91.63 & 84.89 & 92.64 \\ 
        \midrule
        \multirow{3}{*}{Technological} & Bicycle trips (min. 7)  & 725 & 53,118 & 0.1 & 59.53 & 2.75 & 4.62 \\
        & U.S. Airports 2006 & 1,075 & 18,906 & 1.64e-2 & 27.59 & 18.99 & 68.83 \\
        & Water pipes & 1,836 & 2,351 & 6.98e-4 & 99.62 & 95.83 & 96.20 \\
        \bottomrule
    \end{tabular}
    \label{tab:summary-results}
\end{table}

\subsection{Giraffe socialization.}

Evidence suggests that giraffes have complex social structures, with females having social preferences and suggestive that adult giraffes have friendships beyond only mother-child interactions \cite{Bashaw:2007}.
We analyze a network of social interaction of captive giraffes at the San Diego Zoo's Wild Animal Park.
The original observational study included 6 adult female Rothschild’s giraffe (\textit{Giraffa camelopardalis}) housed in a single herd.
%
%
In the study, they were observed 5 mornings a week for a total of 300 days, and the behavior of each subject was recorded for a 20-min focal sample in random order.
Data on nearest neighbor and proximity (measured at 2 neck lengths) were collected at 1-min intervals for the focal subject.
Affiliative social interactions involving the focal subject were recorded and included: approach, necking, head rub, bumping, social exam, muzzle, co-feed, and sentinel (details in \cite{Bashaw:2007}).
In total, 600 hours of observation time and 2,748 affiliative interactions were observed.

\begin{figure}[ht!]
    \centering
    \includegraphics[width=\textwidth]{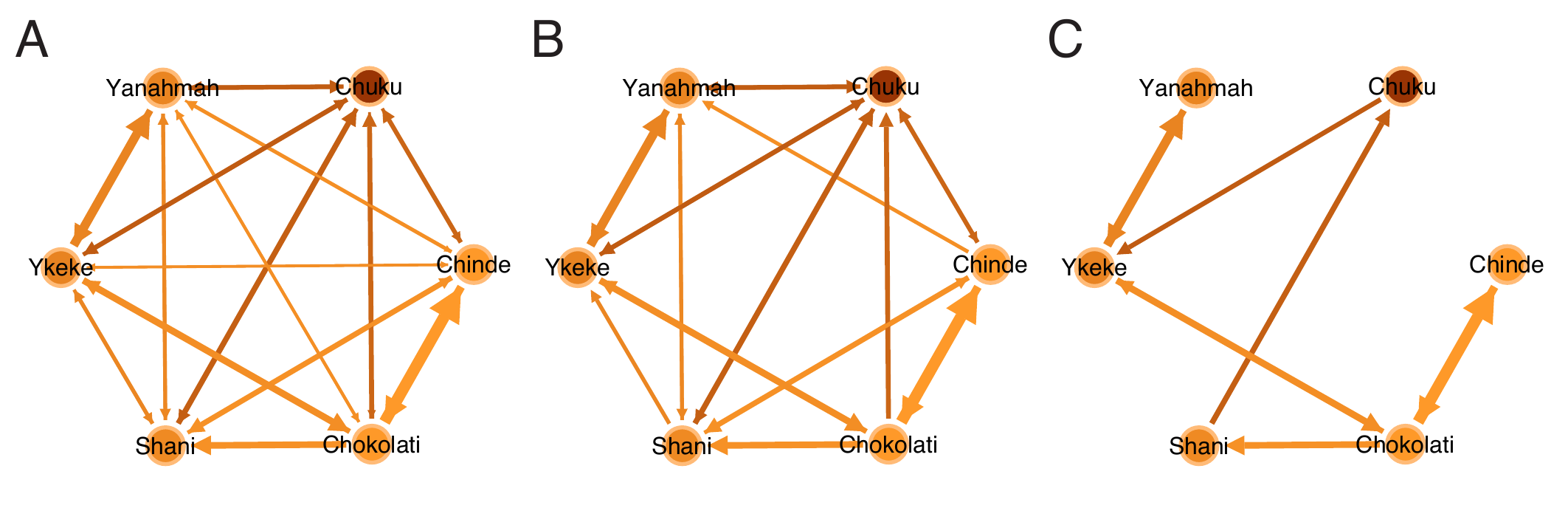}
    \caption[Giraffe socialization network]{
        Giraffe socialization network in the San Diego Zoo \cite{Bashaw:2007}.
        \textbf{(A)} Directed distance graph;
        \textbf{(B)} Metric backbone subgraph; and
        \textbf{(C)} Ultrametric backbone subgraph.
        The original distance graph contains 30 edges, while the metric backbone contains 23 ($76.7\%$) and the ultrametric backbone only 9 ($30\%$) edges.
        Plotted with Gephi \cite{Bastian:2009}.
    }
    \label{fig:giraffe}
\end{figure}

In the social network directed edge weights represent the frequency in which giraffe $x_i$ interacts with giraffe $x_j$ as a measure of similarity $p_{ij}$ (see \cref{fig:giraffe}{A}).
This is a small network containing only 6 nodes and fully connected with 30 directed edges (density $\delta = 1.0$).
The metric backbone consists of 23 edges ($\tau^m=76.7\%$) and the ultrametric of only 9 ($\tau^u=30\%$) edges.
Interestingly, the metric backbone completely removes the edge between giraffes \textit{Yanahmah} and \textit{Chokolati}, both the oldest giraffes in the herd.
In the metric backbone the mother-daughter relationships are also kept between \textit{Yanahman}-\textit{Ykeke} and \textit{Chokolati}-\textit{Chinde}.
In other words, and as previously noted for human contact networks \cite{Correia:2022:epibackbone}, the backbone preserves the hierarchical structure of social networks.

\subsection{London bike-sharing trips.}

The SARS-Cov-2 pandemic caused unprecedented shifts in urban mobility with bike-share systems having a significant increased in demand in several major capitals \cite{SongJie:2022,QiumengLi:2022}.
We analyze the City of London's bike-sharing system, available through the Transport for London Open Data API\footnote{\url{https://api.tfl.gov.uk/}} and previously analyzed in Munoz-Mendez \textit{et al} \cite{MunozMendez:2018}.
Data contains records for each unique bicycle and their rental transactions, including timestamped information on which bike-sharing station it was picked up and then returned in a network of 770 stations through the city.
A month's worth of bike-sharing transactions is analyzed, from June to July 2014.
Transactions that started or ended in a repair station, as well as stations with too few transactions, were discarded. This means we only included stations that accounted for 75\% of all transactions (i.e., a minimum of 7 monthly trips per station), which in turn resulted in 726 bike-sharing stations and 948,339 bike-sharing transactions.

In this network a node represents a bike-sharing station, $x_i$, and edges are weighted by the average trip duration between stations as a directed distance, $d_{ij}$.
This network has 725 nodes and 53,118 nodes (density $\delta=0.1$).
The metric and ultrametric backbones consist of $\tau^m = 59.53\%$ and $\tau^u = 2.75\%$, respectively, of the directed network.
Along with the co-morbidity risk network, the bike sharing network has one of the largest differences in the sizes of the metric to the ultrametric backbone ($\tau^u / \tau^m = 4.6\%$).
This means that a directed attack on the metric backbone will have a small impact on ultrametric backbone and thus in the distribution of shortest paths \cite{Simas:2021}.
In other words, the network of the bike-sharing system for the City of London is very robust to directed attacks, translated to the possible closure of bike-sharing stations or street changes that cyclists use.


\subsection{U.S. airport transportation.}

\begin{figure}[ht!]
     \centering
     \includegraphics[width=\textwidth]{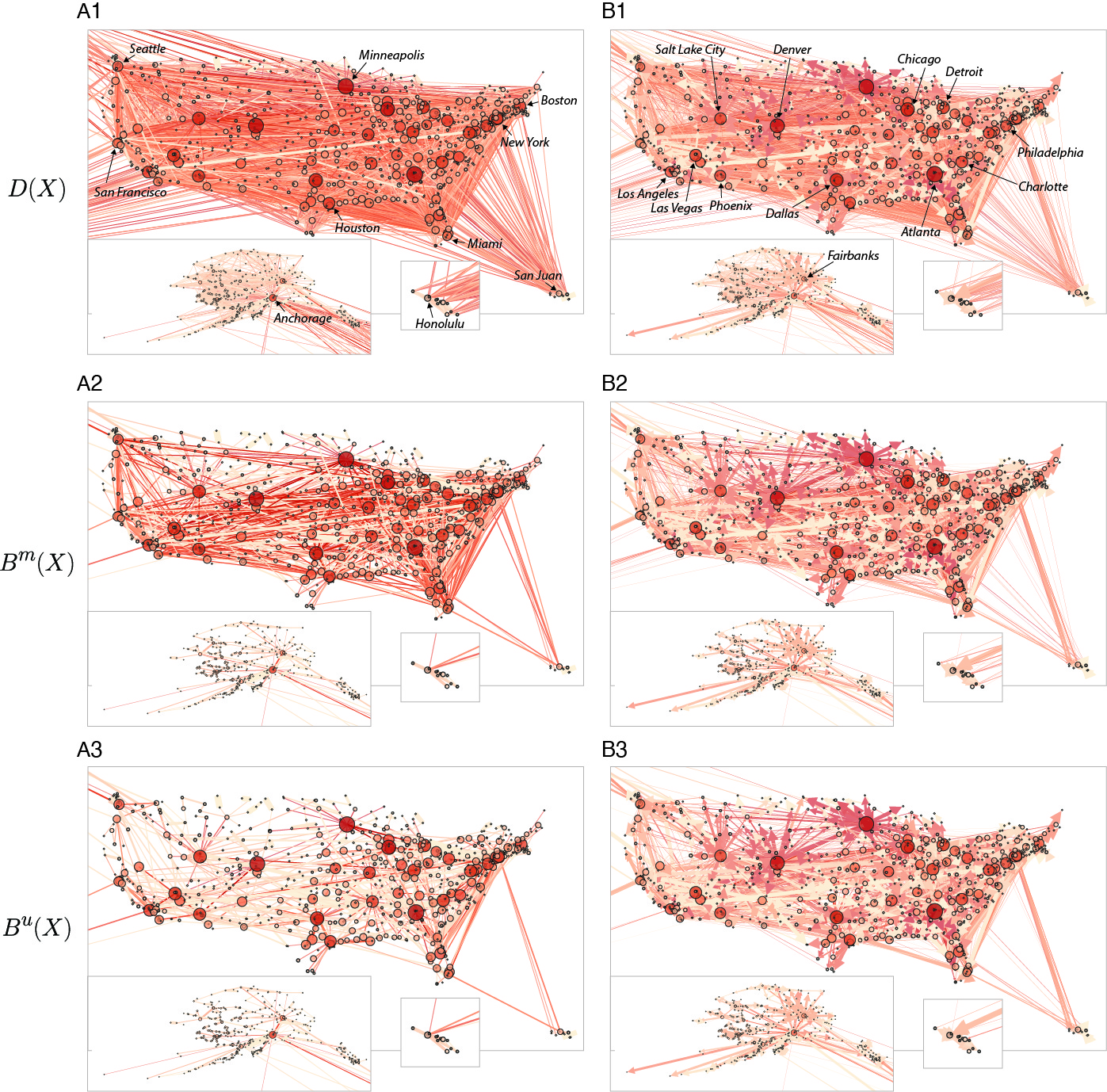}
     \caption[U.S. Airports]{
         Domestic nonstop segment of the U.S. airport transportation system \cite{Serrano:2009}.
         \textbf{(A1)} Undirected distance graph with its respective 
         \textbf{(A2)} metric, and
         \textbf{(A3)} ultrametric backbone subgraphs \cite{Simas:2021}.
         \textbf{(B1)} Directed distance graph with its respective 
         \textbf{(B2)} metric, and
         \textbf{(B3)} ultrametric backbone subgraphs.
         The original directed (undirected) distance graph contains 18,906 (11,973) edges. From those, $27.59\%$ ($16.14\%$) are in the metric backbone, and $18.99\%$ ($8.98\%$) in the ultrametric backbone.
         The difference in number of edges between the undirected and directed representation comes from the fact that 5,040 ($26.65\%$) of all flights are only in one direction.
         Network plotted with Gephi \cite{Bastian:2009}.
     }
     \label{fig:bikeshare}
\end{figure}

This network is the domestic nonstop segment of the U.S. airport transportation system for the year 2006, retrieved from \url{http://www.transtats.bts.gov}.
Each node is an airport, and edge weights are the normalized number of passengers traveling between two airport-nodes.
This network was analyzed in Simas \textit{et al.} \cite{Simas:2021} and is a reconstruction of the one used by Serrano \textit{et al.} \cite{Serrano:2009}.
Differently from previous work, however, here we consider directionality in the flow of passengers as 5,040 (approximately 27\%) of all flights are only in one direction
In other words, flight routes may include stops in multiple airports from initial to final destination, and not necessarily contain a direct return to the initial departure airport.
Airports in the American Samoa, Guam, Northern Marianas, and Trust Territories of the Pacific Islands have been removed from the analysis.
This is a large but relatively sparse network with 1,075 nodes and 18,906 edges (density $\delta = $1.64e-2).
%
%
%
The relative size of the metric and ultrametric backbone are $\tau^m = 27.59\%$ and $\tau^u = 18.99\%$, respectively, (see \cref{tab:summary-results}).
%
%
%


\section{Discussion}


Directionality and strength of interactions are relevant properties of real complex networks.
The structure of such networks can be reduced in a principled manner, while preserving the entire distribution of shortest paths (for a given length measure $g$), with the computation of the distance backbone.

In the nine networks we analyzed, we found that the size of the metric backbone ranges from 27.59\% to 99.6\% – three networks have metric backbones above 92\% of the distance graph.
Ultrametric backbones range from 2.17\% to 99.5\%, with two networks having ultrametric backbones above 95.8\%.
In contrast, for undirected graphs studied in Simas \textit{et al.} \cite{Simas:2021} the metric (ultrametric) backbones range from 1.75\% to 83.59\% (0.2\% to 78.45\%), which shows a substantial increase in the size of backbones due to directionality. 
A direct comparison can be made for the U.S. airports network. Its undirected representation has a relative size of the metric and ultrametric backbone of $\tau^m = 16.14\%$ and $\tau^u = 8.98\%$, respectively \cite{Simas:2021}. Here, we found that the relative size of the metric and ultrametric backbone are $\tau^m = 27.59\%$ and $\tau^u = 18.99\%$, respectively (see \cref{tab:summary-results}).
This increase is likely due to the fact that the closure for directed graphs does not lead to a complete graph---unlike what happens to connected undirected graphs.
In other words, having many connections in only one direction (approximately 27\% in this case) can make them necessary for shortest paths irrespective of the edge weight, which emphasized the importance of directionality when studying real-world networks.
The large difference between the size of backbones in directed and undirected graphs warrants future studies of the effect of directionality vis a vis various topological parameters.


The metric and ultrametric backbones of the networks we analyzed (\cref{tab:summary-results}) exemplify networks which are robust to random edge removal, as is the case of the comorbidity risk and bicycle trips networks, for having a smaller backbone (small $\tau^g$).
On the other hand, the species-species interaction network and water pipes networks have a large $\tau^g$ and little redundancy. That is, the backbone is most of the network, suggesting that they mostly contain necessary interaction information, or were perhaps optimized to minimize the cost of implementing redundant edges, being susceptible to random edge removal or failure.
In the case of the water pipe network, little redundancy is expected because its distance weights represent an actual physical distance between nodes, which must conform to a naturally metric topology.
Thus, it is an expected result that its metric backbone is almost the entire distance graph ($99.6\%$). 
This highlights the fact that semi-metric (and semi-triangular) behavior can only occur in high-dimensional spaces \cite{Simas:2021}.
In contrast, the  metric backbone of  the passenger traffic between U.S. airports is only $27.59\%$, making its shortest path distribution very robust to random attacks, as the odds of randomly removing semi-metric edges are much higher than removing metric ones that contribute to the backbone.
The precise impact in the shortest path distribution for those networks requires the computation of edge distortion \cite{Simas:2012,Simas:2021} and is left for future work.

\section{Conclusion}

We introduced directionality to study shortest-path redundancy in \emph{weighted directed} graphs via a novel directed distance backbone subgraph, the computation of which we showed to be feasible.
This consideration brings improvement over other sparsification methods that considers only undirected networks \cite{Mercier:2022,Simas:2021} or that treat incoming and outgoing edges independently \cite{Serrano:2009}.
%
%
We focused on the metric (where $g \equiv +$) and the ultrametric (where $g \equiv max$) backbones, but the methodology is applicable for any length measure $g$, allowing other backbones to be considered in the future.

We applied the methodology to study redundancy of a variety of real-world weighted directed graphs modeling biomedical, social, and technological systems.
The size of the metric (ultrametric) backbone ranges from 27\% to 99\% (2\% to 99\%), but is typically much smaller than the original distance graph.
However, the size of the directed backbones observed are larger than the undirected backbones previously reported, emphasizing the difference in shortest-path robustness for the two different classes of graphs.
The comparison using the same underlying  U.S. airports network is particularly illuminating.
We found that both the metric and the ultrametric backbone for the directed graph are larger than the ones for the undirected version---$71\%$ and $112\%$, respectively.
Thus, asymmetric airline seat capacity between cities (27\% of all connections exist only in one direction) has a large impact on shortest paths between them.
This exemplifies the importance of our contribution in the study of distance backbones for directed networks, which will lead to a study with additional networks in the future.

The methodology further allows us to infer the robustness of shortest path distributions to random attack, via the relative size of the metric and ultrametric backbones. This can aid the design of more resilient social and technological systems or the identification of key evolutionary properties in biomedical systems.
We are confident that the study of directed distance backbones can help the understanding and control of a variety of complex multivariate systems where both strength and directionality of interactions is key.

\bibliographystyle{splncs03}
\bibliography{reference}
\vfill


\end{document}